\documentclass[11pt]{llncs}
\usepackage[letterpaper,hmargin=1in,vmargin=1.25in]{geometry}
\usepackage{color}
\usepackage{epsf}
\usepackage{url}
\usepackage{amssymb,amsmath}
\usepackage{algorithm}
\newcommand{\anomsg}{\textsf{anonymous message transmission}}
\newcommand{\Anomsg}{\textsf{Anonymous Message Transmission}}

\newcommand{\authc}{\textsf{authenticate}}
\newcommand{\authd}{\textsf{decode}}

\newcommand{\qt}{\textsf{fail-safe quantum teleportation}}

\newcommand{\squash}[1]{\raisebox{0.04ex}[0pt][0pt]{\small$\textstyle #1$}}
\newcommand{\oosrt}{\squash{\frac{1}{\sqrt{2}}}}

\floatname{algorithm}{Protocol}

\def\ket#1{{\lvert}#1\rangle}
\providecommand{\abs}[1]{\lvert#1\rvert}

\begin{document}

\title{Anonymous quantum communication\\
\small{(15 June 2007)}}
\author{Gilles Brassard\inst{1}, Anne Broadbent\inst{1}, Joseph Fitzsimons\inst{2}, S\'{e}bastien Gambs\inst{1}  and Alain Tapp\inst{1}}
\institute{Universit\'{e} de Montr\'{e}al\\
D\'{e}partement d'informatique et de recherche op\'{e}rationnelle\\
C.P. 6128, Succ.\ Centre-Ville, Montr\'{e}al (Qu\'ebec), H3C 3J7 \textsc{Canada}\\
\email{\{brassard,broadbea,gambsseb,tappa\}\textnormal{@}iro.umontreal.ca}\\
\and University of Oxford\\ Department of Materials\\ Parks Road, Oxford, OX1 3PH \textsc{United Kingdom}\\
\email{joe.fitzsimons@materials.ox.ac.uk}}

\hyphenation{avenues}
\hyphenation{research}
\hyphenation{interact}
\hyphenation{machine}
\hyphenation{machines}
\hyphenation{giving}
\hyphenation{paper}
\hyphenation{encoun-ter}
\hyphenation{encoun-ters}
\hyphenation{expe-ri-ence}
\hyphenation{analyse}
\hyphenation{analysis}
\hyphenation{remains}
\hyphenation{logical}
\hyphenation{cannot}

\clubpenalty 10000
\widowpenalty 10000

\maketitle

\begin{abstract}
We present the first protocol for the anonymous transmission of a
quantum state that is information-theoretically secure against an
active adversary, without any assumption on the number of corrupt
participants. The anonymity of the sender and receiver is perfectly
preserved, and the privacy of the quantum state is protected except
with exponentially small probability. Even though a single corrupt
participant can cause the protocol to abort, the quantum state can
only  be destroyed with exponentially small probability: if~the
protocol succeeds, the state is transferred to the receiver and
otherwise it remains in the
hands of the sender (provided the receiver is honest).\\[1ex] \textbf{Keywords:} quantum cryptography, multiparty computation,
anonymity, dining cryptographers.
\end{abstract}

\section{Introduction}

In David Chaum's classic dining cryptographers
scenario~\cite{Chaum88}, a group of cryptographers is having dinner
at a restaurant and it is the case that either one of them has
anonymously paid the dinner bill or the NSA has paid. The task that
the cryptographers wish to accomplish is to find out which of the
two cases occurred, without revealing any additional information.
The security of Chaum's protocol does not rely on any computational
assumption, but only on the cryptographers having access to pairwise
private channels and to a broadcast channel. A simple extension to
this protocol allows a single participant, say Alice, to broadcast a
message to all the other participants in such a way that Alice's
identity is information-theoretically protected.

But what if Alice wishes to send a private message to Bob (who is
also sitting at the dinner table), while ensuring the anonymity of
both herself and of Bob? This task is called~\emph{anonymous message
transmission}. As an instance of multiparty secure computation, such
a protocol can be accomplished, assuming pairwise private channels
and a broadcast channel, as long as a majority of participants are
honest~\cite{RB89}. Recently, two of us~\cite{BT07} have given a
protocol that requires pairwise private channels and  a broadcast
channel, and accomplishes anonymous message transmission
\emph{without} any assumption on the number of honest participants
(the protocol, however, allows even a single corrupt participant to
cause an abort).

Our main contribution is to give the first information-theoretically
secure protocol for  \emph{quantum} anonymous transmission that
tolerates any number of corrupt participants. That is, our protocol
allows Alice to send a quantum message to Bob such that both Alice
and Bob remain anonymous (no participant learns the identity of
Alice---even if Bob is corrupt---and the identity of Bob remains
known only to Alice), and the quantum message remains private
(nothing about it leaks to participants other than Bob, unless of
course Bob is corrupt). The anonymity of the sender and receiver, as
well as the privacy of the quantum message, are perfect, regardless
of the behaviour of cheating parties, with no need to rely on any
assumptions other than the availability of a classical broadcast
channel as well as private authenticated quantum channels between
each pair of participants. Our protocol has features similar to the
anonymous (classical) message transmission protocol mentioned above:
we can tolerate an arbitrary number of corrupt participants, but any
single corrupt participant can cause the protocol to abort. However,
no private information can be obtained by making the protocol abort.

Since Alice sends quantum information, we need to address a concern
that did not exist in the context of classical anonymous message
transmission: the state to be transmitted should never be destroyed
\emph{even if the protocol aborts} (unless the receiver is corrupt,
since in that case he can follow honestly the protocol until the
very end, and then destroy the successfully transmitted message!).
Because of the no-cloning theorem~\cite{WZ82}, the sender cannot
keep a backup copy of the message before entering the protocol.
Nevertheless, we accomplish this safeguard as part of the main
protocol with a simple and novel notion called \emph{fail-safe
teleportation}.

\subsection{Anonymity}

\emph{Anonymity} is a basic cryptographic property whose goal is to
hide the identity of the sender or receiver of a message (or~both).
It is different from, but often complementary to \emph{privacy},
which ensures the confidentiality of a message. Examples of
anonymous tasks include sending an anonymous letter to one's love,
using an email account with a pseudonym, accessing a web page
through a trusted identity proxy server or blind reviewing of a
conference paper. Three approaches to classical anonymity are
generally considered. The first one requires the help of a trusted
third party that forwards messages between participants without
revealing the identity of the senders.
Anonymizers~\cite{Boyan97,GGKMA99} belong to this class. The second
approach uses chains of untrusted servers that randomize the
ordering of messages. This reordering prevents an outside observer
from linking the sender and the receiver of a particular message.
The privacy of messages is generally assured by a public-key
cryptosystem. Chaum's MixNets~\cite{Chaum81} are an instance of
techniques using this approach. The third and last approach offers
information-theoretic security, assuming resources such as a
broadcast channel and pairwise private channels. Chaum's dining
cryptographers protocol~\cite{Chaum88} is the archetypical example
of a protocol in this category.

\subsection{Model}

In our model,  we  suppose that each pair of participants shares a
\emph{private authenticated quantum channel}, which means that a
participant can send an authenticated private message (quantum or
classical) to any other participant. Such a channel can be
implemented if the participants share pairwise quantum channels as
well as classical secret keys.  An extra tool is given to the
participants under the form of a (classical) \emph{broadcast channel}. This
channel guarantees that all participants receive the same message
from a publicly known sender, and that the message is not modified
while in transit.

Two security models are generally considered in secure multiparty
computation: \emph{honest-but-curious} and \emph{malicious}. In the
honest-but-curious model (also called \emph{semi-honest}), the
participants are assumed to follow
 the protocol (thus being honest) but at the same time
record all the information they have seen during its execution (thus
being curious). In this model, a protocol is said to be secure
against a \emph{collusion} of participants if, by pooling their
data, these participants cannot learn more information than from
their inputs and the output of the protocol alone. In the malicious
model, participants may actively cheat and deviate from the original
prescription of the protocol. Cheaters can for instance try to learn
information about the input of honest participants or tamper with
the output of the protocol. Formal definitions can be found in
Chapter~7 of~\cite{Goldreich04}. Both these models are neatly
encapsulated by considering a central entity called an
\emph{adversary}, which controls some of the participants, rendering
them \emph{corrupt}. The adversary is \emph{passive} if the corrupt
participants are honest-but-curious, and \emph{active} if the
corrupt participants are malicious. In this paper, we consider the
case of an active adversary that chooses the set of corrupt
participants before the execution of the protocol.

In the scenario that we consider, within a group of~$n$
participants, the anonymous sender communicates a private quantum
message to an anonymous receiver. The sender is unknown to all
participants and the receiver is unknown to all participants except
to the sender. We give the following formal definitions, adapted
from~\cite{chrweh05}:

\begin{definition}[Sender anonymity]\label{def:send-anon}
A protocol achieves \emph{sender anonymity} if at the end of the
protocol, the probability that an adversary controlling any
number~$t$ of participants (excluding the sender)  can  correctly
guess the identity of the sender is at most $\frac{1}{n-t}$. If the
sender is corrupt, then the protocol vacuously achieves sender
anonymity.
\end{definition}

\begin{definition}[Receiver anonymity]\label{def:rec-anon}
A protocol achieves \emph{receiver anonymity} if at the end of the
protocol, the probability that an adversary controlling any
number~$t$ of participants (excluding the sender and receiver) can
correctly guess  the identity of the receiver is at most
$\frac{1}{n-t}$. If either the sender or receiver is corrupt, then
the protocol vacuously achieves receiver anonymity.
\end{definition}

The intuition behind these definitions is that a protocol for
completing an anonymous task should not reveal any information about
the identity of the sender or of the receiver. If this property is
verified, the best an adversary can do at the end of the protocol is
to guess at random their identities. If the sender is corrupt, then
there is no sender anonymity to preserve; a similar observation
applies to receiver anonymity. Note however that sender anonymity
requires that no adversary can learn the identity of the sender,
\emph{even if the receiver is corrupt}.

In what follows,  we are only interested in protocols that are
unconditionally secure in the information-theoretic sense for the
purpose of achieving sender and receiver anonymity. We place no
limit on the number of corrupt participants. It is therefore not
surprising  that the protocol could abort if even a single corrupt
participant deviates from the prescribed protocol. Even if the
protocol aborts, sender and receiver anonymity, and message privacy
are never compromised. Note that if we had some sort of guarantee
that a strict majority of participants is honest, then anonymous
quantum message transmission could be implemented as a special case
of quantum secure multiparty computation~\cite{BCGHS06}.

\subsection{Anonymity in the quantum world}

The first  protocol based on quantum mechanics that allows the
anonymous communication of \emph{classical} information was proposed
by P.~Oscar Boykin~\cite{Boykin02}. In the case of a \emph{quantum}
message, Matthias Christandl and Stephanie Wehner were first to
define the concept of \emph{anonymous quantum message transmission}
and to give an explicit protocol for solving this
task~\cite{Wehner04,chrweh05}, but under the \emph{deus ex machina}
assumption that the $n$ participants share ahead of time entangled
state \mbox{$\ket{+_n}=\oosrt\ket{0^n} + \oosrt\ket{1^n}$}.
(No~mechanism is proposed to verify the validity of that state.)
Under that assumption, their protocol is information-theoretically
secure in terms of sender and receiver anonymity, but malicious
participants can alter the transmitted state in a way that will not
be detected by the honest participants.

One key notion introduced in the paper of Christandl and Wehner is
that of \emph{anonymous entanglement}. Starting with the assumed
$n$-party entangled state $\ket{+_n}$, the sender and the receiver
end up sharing a two-party entangled state $\ket{+_2}$, better known
as Bell State \mbox{$\ket{\Phi^+} = \oosrt\ket{00} +
\oosrt\ket{11}$}, provided the other parties follow the protocol
honestly. This entanglement is \emph{anonymous} because the sender
has chosen with which other party (the receiver) he shares it, but
the receiver has no information concerning the party with which he
is entangled. Moreover, the other parties have no information
concerning who are the two entangled parties (assuming the entangled
parties are not corrupt).

A first attempt to accomplish quantum message transmission in the
presence of an unlimited number of corrupt participants
\emph{without} assuming that a trusted state~$\ket{+_n}$ is shared
between the participants before the onset of the protocol was made
by Jan Bouda and Josef \v{S}projcar~\cite{BS05}, but in a
public-receiver model (the sender is anonymous but the receiver is
public). The creation and distribution of a $\ket{+_n}$ state is an
important part of their protocol. {From} there, they attempt to
establish semi-anonymous entanglement (the identity of one of the
entangled parties, the receiver, is public). However, careful
analysis reveals that an active adversary can proceed in such a way
that the probability that the protocol aborts becomes correlated
with the identity of the sender, thus compromising his anonymity.
If~the protocol requires the receiver to stay quiet in order not to
reveal whether or not the protocol has succeeded, it is true that
the anonymity of the sender is preserved. However, this is very
different from the model usually considered in secure multiparty
computation, in which all the participants learn at the end of the
protocol whether or not it has succeeded. More importantly, this
approach makes it impossible to preserve the identity of the sender
whenever the receiver is corrupt.

Our~own protocol is also based on the establishment of anonymous
entanglement between the sender and the receiver. However, compared
to the protocol of Christandl and Wehner, we do not need to assume
an \emph{a~priori} shared $\ket{+_n}$ state and no malicious attempt
at corrupting the intended final $\ket{\Phi^+}$ state between the
sender and the receiver can succeed (except with exponentially small
probability) without causing an abort. It~follows that the intended
state will be transmitted faithfully unless the protocol aborts, in
which case it will end up intact at the sender's by virtue of
fail-safe teleportation (unless the receiver is corrupt). Compared
with the protocol of Bouda and \v{S}projcar, our receiver is
anonymous and the identity of the sender and the receiver cannot be
correlated with the probability that the protocol aborts, allowing
us to achieve perfectly sender and receiver anonymity according to
Definitions~\ref{def:send-anon} and~\ref{def:rec-anon}.

\section{Toolbox}

We now survey the classical and quantum tools that are used in our
main protocol. Two of us recently developed several classical secure
multiparty protocols~\cite{BT07}; we present below some of the
relevant results, which will be used in the next section. All
protocols assume pairwise authentic private classical channels and a
broadcast channel. They offer information-theoretic security and
have polynomial complexity in the number of participants as well as
in a safety parameter and, in the case of
Theorem~\ref{theorem:anonymous message}, in the number of bits in
the transmitted message. In~all cases, the expression
``exponentially close to~1'' or ``exponentially small'' means
``exponentially in the safety parameter''. We~also review a key
result from~\cite{BCGST02}.

\begin{theorem}[\textsf{Logical~OR}--\cite{BT07}] \label{thm:OR}
There exists a  secure multiparty protocol to compute the
\textsf{logical~OR} of the participants' input bits (one bit per
participant). Misbehaving participants cannot cause the protocol to
abort. (Any refusal to participate when expected will cause the
output to be~1.) The~correct answer is computed with probability
exponentially close to~1. The only information an active adversary
can learn through the protocol is if at least one honest participant
has input~1. No~information about the number of such participants or
their identity is revealed.
\end{theorem}

\begin{theorem}[\textsf{Collision Detection}--\cite{BT07}] \label{thm:collision}
There exists a \textsf{collision detection} protocol in which each
participant inputs a bit.  Let~$r$ denote the number of $1$s among
these input bits. The protocol has three possible outcomes
corresponding to whether \mbox{$r=0$}, \mbox{$r=1$} or~\mbox{$r \geq
2$}. If~all participants are honest, the correct value is computed
with probability exponentially close to 1. No~participant can make
the protocol abort, and an adversary cannot learn more than it could
have learned by assigning to all corrupt participants the input~$0$
and letting them follow the protocol faithfully. A~single corrupt
participant can cause the output corresponding to~$r \geq 2$
regardless of the other inputs (even if all the other inputs
are~$0$). Also, it is possible for a corrupt participant to set his
input to~$0$ if all other participants have input~$0$ (producing an
\mbox{$r=0$} output) and to~$1$ otherwise (producing an \mbox{$r
\geq 2$} output). No~other form of cheating is possible.
\end{theorem}

Although the \textsf{collision detection} protocol outlined above
may look rather imperfect, it is actually just as useful as the
ideal protocol for our purpose.

\begin{theorem}[\textsf{Notification}--\cite{BT07}] \label{thm:notification}
There exists a \textsf{notification} protocol in which participants
can notify other participants of their choosing. Each player's
output is one private bit specifying if he has been notified at
least once; this value is correctly computed with probability
exponentially close to 1. This is the only information accessible
through the protocol even in the case of an active adversary.
\end{theorem}

According to~\cite{BT07}, it is possible in general to invoke the
\textsf{notification} protocol even if multiple senders want to
notify several receivers. However, in the specific context of our
use of this protocol for the purpose of anonymous quantum message
transmission, we forbid any honest participant to engage in the
above \textsf{notification} protocol without having previously
caused output ``\mbox{$r=1$}'' in the \textsf{collision detection}
protocol (Theorem~\ref{thm:collision}). Similarly, no~honest
participant $S$ will ever engage in the {\anomsg} protocol below
unless he has initially caused output ``\mbox{$r=1$}'' in the
\textsf{collision detection} protocol \emph{and} has
\textsf{notified} a single other participant~$R$.

\begin{theorem}[\Anomsg{}--\cite{BT07}] \label{theorem:anonymous message}
There exists an {\anomsg} protocol in which a sender~$S$ can
transmit a classical message to a receiver~$R$ such that the
anonymity of~$S$ and~$R$ and the privacy of the message are perfect
even in the presence of an active adversary.
If~all participants are honest then the message is transmitted
perfectly. Any attempt by a corrupt participant  to modify the
message will cause the protocol to abort, except with exponentially
small probability.
\end{theorem}

In 2002, Howard Barnum, Claude Cr\'epeau, Daniel Gottesman and Alain
Tapp presented a non-interactive scheme for the authentication of
quantum messages~\cite{BCGST02}. The protocol also encrypts the
quantum state to be transmitted and is information-theoretically
secure.

\begin{theorem}[\textsf{Quantum Authentication}--\cite{BCGST02}]
There exists an information-theoretically secure quantum
authentication scheme to authenticate an arbitrary quantum message
$\ket{\psi}$ of length~$m$ with an encoding circuit (called \authc)
and a decoding circuit (called \authd) of size polynomial in~$m$,
which uses a random private key of length~$2m + 2s+1$ and has
authenticated message of length~$m+s$. Let~$p$ the probability that
the message is accepted. If the message is accepted then let~$q$ be
the probability of obtaining outcome~$\ket{\psi}$ when measuring in
a basis containing~$\ket{\psi}$. If~the authenticated message is not
modified, then~$p=q=1$. Otherwise, $p q + (1-p)> 1- \frac{m+s}{s(2^s
+ 1)}$. The protocol also perfectly preserves the privacy of the
transmitted message.
\end{theorem}

\section{Protocol for anonymous quantum message transmission}

In this section, we describe and analyse our protocol for anonymous
quantum message transmission. Our protocol allows an anonymous
sender~$S$ to transmit an~$m$-qubit message~$\ket{\psi}$ to an
anonymous receiver~$R$\@. We assume a broadcast channel as well as
an information-theoretically secure private and authenticated
quantum channel between each pair of participants (which can also be
used, of course, to transmit classical information). Our protocol
perfectly preserves the anonymity of the sender and receiver, as
well as the privacy of the message. The security proof for the
protocol makes no assumption on the number of corrupt participants.
It~is therefore not surprising that a single participant can make
the protocol abort. However, if the sender and the receiver are
honest, the quantum message to be transmitted will only be lost with
exponentially small probability.

Here is an informal description of the protocol. In the first step,
the purely classical \textsf{collision detection} protocol of
Theorem~\ref{thm:collision} is performed to establish that exactly
one participant wants to send an anonymous quantum message. If~this
is not the case, the protocol aborts. In~case it is found that more
than one participant wants to speak, one might imagine alternative
scenarios such as asking each one of them to decide at random
whether or not to skip their turn and trying again the
\textsf{collision detection} protocol until a single-sender
occurrence occurs. This will reveal information on the number of
honest would-be senders and may take too many trials if there are
too many of them, so that more sophisticated solutions might need to
be considered. (We~do not elaborate on this issue for simplicity.)

In the next two steps, the participants collaborate to establish
multiple instances of a shared state~$\ket{+_n} = \oosrt\ket{0^n} +
\oosrt\ket{1^n}$. Then, the sender designates a receiver by use of
the \textsf{notification} protocol (Theorem~\ref{thm:notification}).
If~honest, the receiver will act differently from the other
participants, but in a way that is indistinguishable, so that
his anonymity is preserved. The shared instances of~$\ket{+_n}$
are then used to create anonymous entanglement between the sender
and the receiver. However, the anonymous entanglement could be
imperfect if other participants misbehave. For this reason, the
sender then creates a sufficient number of instances of Bell
state~$\ket{\Phi^+}$. The possibly imperfect anonymous entanglement
is used to teleport~\cite{BBCJPW93} an authenticated version of half
of each~$\ket{\Phi^+}$. If~this first teleportation is successful,
the sender uses this newly established perfect anonymous
entanglement to teleport the quantum message itself. Our \qt\
protocol ensures that unless the receiver is corrupt, the quantum
message is never destroyed, except with exponentially small
probability: either it is safely transmitted to the receiver, or it
comes back intact at the sender's.

In more detail, all classical communication from the sender to the
receiver is performed anonymously using the \anomsg~protocol
(Theorem~\ref{theorem:anonymous message}). To~create anonymous
entanglement, all participants must be involved. One participant
(who is chosen arbitrarily, for instance the first participant in
lexicographic order) creates a state~$\ket{+_n}$ and distributes one
qubit to each participant, keeping one for himself. Of~course, this
participant could be corrupt, so that there is no guarantee that a
proper $\ket{+_n}$ has been distributed. Moreover, a corrupt
distributor could send different states to different honest
participants, in the hope that the future evolution of the protocol
may depend on who is the sender and who is the receiver. Foiling
this threat constitutes a key contribution of our protocol. For~this
reason, all participants \emph{verify} this state \emph{without}
destroying it in the next step. If~the verification succeeds, the
state shared amongst all participants is guaranteed to be invariant
under permutation of the honest participants
(Lemma~\ref{lemma:symmetry}), even though it could still not be a
genuine $\ket{+_n}$ state. This ensures sender and receiver
anonymity. Furthermore, the behaviour of the state~$\ket{+_n}$, when
measured by all but two parties in the Hadamard basis, ensures
correctness (unless is aborts) as shown in Theorems~\ref{thm:honest}
and~\ref{theorem:fail-safe}.

The full protocol is given as \textbf{Protocol~\ref{protocol}},
where we denote by~$P$ the \emph{conditional phase change} defined
by $P\ket{0} = \ket{0}$ and $P \ket{1} =  -\ket{1}$. Note that if
two participants (such as the sender and the receiver) share an
instance of Bell state \mbox{$\ket{\Phi^-} = \oosrt\ket{00} -
\oosrt\ket{11}$}, a single participant (such as the sender) can
convert this to a~$\ket{\Phi^+}$ by locally applying the~$P$
operation. Note also that such a local operation (performed by the
sender) has no detectable effect that could be measured by the other
participants (in particular the receiver), which ensures that the
anonymity of the sender is not compromised. It is easy to see that
\textbf{Protocol~\ref{protocol}} has polynomial complexity in~$n$
(the number of participants), $s$ (the security parameter) and $m$
(the length of the message).

\renewcommand{\labelenumi}{\textbf{\arabic{enumi}.}}
\renewcommand{\labelenumii}{\arabic{enumi}.\arabic{enumii}}
\renewcommand{\labelenumiii}{\arabic{enumi}.\arabic{enumii}.\arabic{enumiii}}

\begin{algorithm}
\caption{Anonymous quantum message transmission} \label{protocol}

Let~$s$ be the security parameter and $m$ be the length of quantum
message~$\ket{\psi}$. All quantum communication is performed using
the private authenticated quantum channels.

\begin{enumerate}

\item  \label{step:collision} \textbf{Multiple Sender Detection} \begin{enumerate}
\item The \textsf{collision detection} protocol (Theorem~\ref{thm:collision}) is used to determine
if one and only one participant wants to be the sender. If~not, the
protocol aborts.
\end{enumerate}

\item  \label{step:entanglement-distribution} \textbf{Entanglement Distribution}
\begin{enumerate}
\item One arbitrarily designated participant creates~$2m+s$ instances of
the state~$\ket{+_n}$ and sends one qubit of each instance to each participant,
keeping  one qubit of each instance for himself.
\end{enumerate}

\item  \label{step:entanglement-verification} \textbf{Entanglement Verification}\\
For each of the~$2m+s$ instances:
\begin{enumerate}
\item Each participant makes~$n-1$ \emph{pseudo-copies} of his qubit by
applying a control-not with it as the source and a qubit
initialized to~$\ket{0}$ as the target. One such pseudo-copy is sent to
every other participant.
\item \label{step:subspace measurement}Each participant verifies that all the~$n$ qubits in his
possession are in the subspace spanned by~$\{\ket{0^n},
\ket{1^n}\}$.
\item Each participant broadcasts the outcome of the previous step.  If any outcome is negative, the protocol aborts.
\item Each participant \emph{resets} $n-1$ of his qubits to~$\ket{0}$ by performing~$n-1$ control-not
operations. These qubits are discarded and the one remaining is back
to the state distributed at
step~\ref{step:entanglement-distribution}.
\end{enumerate}

\item  \textbf{Receiver Notification} \label{step:designation}

\begin{enumerate}
\item
The participants execute the \textsf{notification} protocol
(Theorem~\ref{thm:notification}) in which only~$S$ notifies a
single~$R$.
\end{enumerate}

\item \textbf{Anonymous Entanglement Generation}
\label{step:anonymous-entanglement}\\
For each of the~$2m+s$ instances:
\begin{enumerate}

\item All participants except~$S$ and~$R$ measure  in the Hadamard
basis the qubit that remains
 from step~\ref{step:entanglement-verification}.
\item Each participant broadcasts the result of his measurement
($S$ and $R$ broadcast two random dummy bits).

\item $S$ computes the parity of all the bits received during the previous
step (except his own and that of~$R$).

\item If the parity is odd, $S$ applies~$P$, the  conditional
phase change, to his remaining qubit \\
(the two qubits shared by~$S$ and~$R$ are now in Bell
state~$\ket{\Phi^+}$). \end{enumerate}

\item  \textbf{Perfect Anonymous Entanglement}
\label{step:perfect-anon-entanglement}

\begin{enumerate}

\item $S$ creates $2m$ instances of Bell state~$\ket{\Phi^+}$. He keeps the first qubit of each pair; let~$\rho$ be
the rest of the pairs.

\item $S$ creates a random classical key~$k$ of length~$4m+2s+1$, and computes
$\rho^\prime= \authc( \rho,k)$.

\item $S$ performs a teleportation measurement on  $\rho^\prime$
using the anonymous $\ket{\Phi^+}$ states generated during
steps~\ref{step:entanglement-distribution}--\ref{step:anonymous-entanglement}.

\item $S$ uses the \anomsg\ protocol (Theorem~\ref{theorem:anonymous message}) to
send~$k$ and the teleportation bits to~$R$.

\item $R$ completes the teleportation and computes $\rho=$
\authd$(\rho^\prime,k)$. \label{substep:authenticate} If the decoding is
successful, $S$ and~$R$ share perfect anonymous entanglement (they
share~$2m$ instances of~$\ket{\Phi^+}$).
\item A \textsf{logical OR} is computed (Theorem~\ref{thm:OR}): all players input~$0$ except~$R$, who inputs~$1$ if the authentication failed and $0$ otherwise.
If~the outcome is~$1$, the protocol aborts.
\end{enumerate}

\item \textbf{Fail-Safe Teleportation}
\label{step:fail-safe-teleport}

\begin{enumerate}
\item $S$ teleports the state~$\ket{\psi}$ to~$R$ using the first~$m$ pairs generated in the previous step.
The teleportation bits are anonymously transmitted to~$R$
(Theorem~\ref{theorem:anonymous message}). If the communication
succeeds, $R$ terminates the teleportation.\item  A \textsf{logical
OR} is performed (Theorem~\ref{thm:OR}): all players input~$0$
except~$R$, who inputs~$1$ if the communication of the teleportation
bits failed. If the outcome is~$0$, the protocol succeeds.
Otherwise, $S$ and~$R$ do the following: \begin{enumerate}
\item  $R$ performs a teleportation measurement using the remaining
perfect anonymous entanglement to teleport back to $S$ the quantum
state resulting from partially failed
step~\ref{step:fail-safe-teleport}.1.  \item All participants
broadcast $2m$ random bits, except~$R$ who broadcasts the
teleportation bits from above. The protocol continues even if one of
the participants refuses to broadcast.

\item $S$ reconstructs~$\ket{\psi}$ from his own teleportation bits from
step~\ref{step:fail-safe-teleport}.1 and $R$'s teleportation bits
received from the broadcast. The protocol aborts.
\end{enumerate}

\end{enumerate}
\end{enumerate}
\end{algorithm}

\begin{theorem}[Correctness]
\label{thm:honest} Assume all participants are honest in
\textbf{Protocol~\ref{protocol}}. If~more than one of them wishes to
be a sender, this will be detected with probability exponentially
close to~1 in the first step. Otherwise, the message is transmitted
perfectly with probability exponentially close to~1, and the
protocol can abort only with exponentially small probability.
\end{theorem}

\begin{proof}

Even if all participants are honest, it is possible for
\textsf{collision detection} or \textsf{notification} to produce an
incorrect output (the \textsf{notification} protocol may also
abort); however, this happens with exponentially small probability.

To ensure correctness of the protocol, we only have to verify
that~$S$ and~$R$ share a sufficient number of proper Bell
states~$\ket{\Phi^+}$ at the end of
step~\ref{step:anonymous-entanglement}. It~is clear that at the end
of step~\ref{step:entanglement-verification}, the participants share
proper instances of state~$\ket{+_n}$ (since we are assuming in this
theorem that they are honest). When~$S$ computes the parity of the
measurement outcomes in step~\ref{step:anonymous-entanglement}, this
corresponds to the parity of the measurement results in the Hadamard
basis of the state~$\ket{+_n}$, where all but two qubits are
measured. If the parity is even, $S$ and~$R$ share~$\ket{\Phi^+}$
and otherwise~$\ket{\Phi^-}$, which is corrected by the sender by
the application of the conditional phase change $P$. \qed
\end{proof}

The following Lemma is necessary in the proof of anonymity and
privacy (Theorem~\ref{thm:corrupt}).

\begin{lemma} \label{lemma:symmetry}
In \textbf{Protocol~\ref{protocol}}, if
step~\ref{step:entanglement-verification} succeeds, then the state
of the system at the end of the step~is:
\begin{equation} \label{equation:symmetry}
\alpha \ket{00\ldots 0}_H\ket{\psi_0}_C + \beta \ket{11\ldots
1}_H\ket{\psi_1}_C \,,
\end{equation}
where~$H$ denotes the honest participants' subsystem, $C$
denotes the  corrupt participants' subsystem, and $\alpha, \beta \in
\mathbb{C}$ are such that~$\abs{\alpha}^2 + \abs{\beta}^2 = 1$.
\end{lemma}
\begin{proof}
In the entanglement verification step, each honest participant sends
a pseudo-copy of his state to every other honest participant.
Therefore, after a single honest participant verifies that his
qubits are in the subspace spanned by~$\{\ket{0^n}, \ket{1^n}\}$, we
are already ensured that if the entanglement verification succeeds,
the state will be of the form given above.\qed \end{proof}

\begin{theorem} [Anonymity and Privacy]
\label{thm:corrupt} In \textbf{Protocol~\ref{protocol}}, regardless
of the number of corrupt participants, the anonymity of sender~$S$
and~receiver $R$ are always perfect. The privacy of the transmitted
message~$\ket{\psi}$ is perfect, except with exponentially small
probability.
\end{theorem}

\begin{proof}
We analyse the protocol step by step in order to prove the
statement.

By virtue of Theorem~\ref{thm:collision}, step~\ref{step:collision}
does not compromise the identity of the sender, and it involves
neither the receiver nor the quantum state to be transmitted.
Steps~\ref{step:entanglement-distribution}
and~\ref{step:entanglement-verification} are done without any
reference to~$S$ or~$R$ and thus cannot compromise their anonymity
either. Furthermore, the state obtained at the end of
step~\ref{step:entanglement-verification} (if~it does not abort)
cannot be specifically correlated with any honest participant even
if some other participants are corrupt. More precisely, by
Lemma~\ref{lemma:symmetry}, the state is \emph{invariant under any
permutation of the honest participants}. This is crucial for the
anonymity and privacy of the rest of the protocol. In particular, it
guarantees that the probability that the protocol aborts does not
depend on the identity of~$S$ or~$R$\@. We prove this below in the
analysis of step~\ref{step:perfect-anon-entanglement}.

The security of step~\ref{step:designation} follows directly from
the unconditional security of the \textsf{notification} protocol
(Theorem~\ref{thm:notification}). However, if~$S$ fails to
notify~$R$ in step~\ref{step:designation} (this happens with
exponentially small probability), an adversary can surreptitiously
take over the role of the honest receiver in the rest of the
protocol without being detected. In~that case, the adversary will
violate the secrecy of the transmitted state, yet without
compromising the sender and receiver anonymity.

In step~\ref{step:anonymous-entanglement}, anonymous entanglement is
generated. No information is revealed to the adversary in this step
since all communication is done by honest participants broadcasting
random bits.

For step~\ref{step:perfect-anon-entanglement}, all communication is
done using the \anomsg~protocol, which is secure according to
Theorem~\ref{theorem:anonymous message}, except in
\textsf{logical~OR} computation at the end, which reveals the
success or failure of the authentication part of the protocol. We
now show that this last substep cannot reveal any information on the
identity of~$S$ or~$R$\@. This is because the success or failure of
the authentication step is uncorrelated to the identity of~$S$
and~$R$: by Lemma~\ref{lemma:symmetry}, as far as the qubits are
concerned, all honest participants are identical under permutation.
Thus the adversary has no strategy that would allow him to determine
any information about the identity of~$S$ or~$R$.

During step~\ref{step:fail-safe-teleport}, all the bits sent
from~$S$ to~$R$ are randomly and uniformly distributed because they
are the classical bits resulting from the teleportation protocol,
therefore they  do not reveal any information about the identity of
$S$. A similar observation about the bits broadcast by~$R$ in the
case that the very last part of the protocol is executed ensures
that~$R$ and~$S$ remain anonymous.

The privacy of the state~$\ket{\psi}$ in the case that~$S$
successfully notified~$R$ in step~\ref{step:designation} (which
happens with probability exponentially close to~1)  is guaranteed by
the basic properties of teleportation. \qed
\end{proof}

\begin{theorem}\label{theorem:fail-safe}
At the end of  \textbf{Protocol 1}, if~$R$ is honest then the
state~$\ket{\psi}$ is either in the possession of~$S$ or~$R$, except
with exponentially small probability. Furthermore, $\ket{\psi}$ can
only stay with $S$ if the protocol has aborted.
\end{theorem}
\begin{proof}
If all participants are honest, then by Theorem~\ref{thm:honest},
the state is in the possession of~$R$ except with exponentially
small probability. Otherwise, the protocol might abort before
step~\ref{step:fail-safe-teleport}, in which case~$S$ still
has~$\ket{\psi}$. If the protocol reaches
step~\ref{step:fail-safe-teleport}, due to the \textsf{quantum
authentication} of step~\ref{step:perfect-anon-entanglement}, $S$
and~$R$ share~$2m$ perfect Bell states~$\ket{\Phi^+}$ (with
probability exponentially close to~$1$), which are used  for
teleportation in step~\ref{step:fail-safe-teleport}. If~the first
step of the fail-safe teleportation fails, then~$S$ no longer
has~$\ket{\psi}$; however, the last three substeps of the protocol
will always succeed and~$S$ will reconstruct~$\ket{\psi}$ (provided
$R$ is honest). Furthermore, it follows from the virtues of
teleportation that if the protocol does not abort, the state is no
longer with~$S$.
 \qed
 \end{proof}

The reason why we specify in Theorem~\ref{theorem:fail-safe}
that~$R$ must be honest is that a corrupt~$R$ can
destroy~$\ket{\psi}$ by simply discarding it after having faithfully
followed the entire protocol. There~remains one subtlety to mention:
a~corrupt~$R$ could behave honestly until the last step. Then, he
would input~$1$ in the \textsf{logical~OR} computation to force $S$
to accept the teleportation back of the state. At~that point, the
corrupt $R$ could teleport back to $S$ a fake state. As~a result,
$S$ would be fooled into thinking he still has custody of the
original quantum state when, in fact, that state is in the hands
of~$R$. (In~general, there will be no way for $S$ to know that this
has happened.)

\section{Conclusion and discussion}

We have presented the first information-theoretically secure
protocol for quantum communication between an anonymous sender and
an anonymous receiver that tolerates an arbitrary number of corrupt
participants. In particular, this means that no adversary can  learn
any information that will
 break the anonymity of the sender or receiver. Our protocol also provides perfect privacy for the quantum message
and ensures that the quantum message is never destroyed, except with
exponentially small probability. The drawback of our protocol is
that any participant can  disrupt the protocol and make it abort.

\section{Acknowledgements}

We are grateful to Patrick Hayden and  Flavien Serge Mani Onana for
insightful discussions. G.\,B. is supported in part by the Natural
Sciences and Engineering Research Council of Canada
(\textsc{Nserc}), the Canada Research Chair program, and the
Canadian Institute for Advanced Research (\textsc{Cifar}). A.\,B. is
supported in part by scholarships from the Canadian Federation of
University Women and the Fonds Qu\'ebecois de Recherche sur la
Nature et les Technologies (\textsc{Fqrnt}). J.\,F. is supported by
a Helmore Award.
S.\,G. is supported by \textsc{Nserc}.
A.\,T. is supported in part by \textsc{Cifar},
\textsc{Fqrnt}, the Mathematics of Information Technology and
Complex Systems Network, and \textsc{Nserc}.
Furthermore, we acknowledge the support of \textsc{Intriq} and QuantumWorks.

\bibliography{anonymous_quantum_communication}
\bibliographystyle{alpha}

\end{document}